\documentclass[letterpaper]{article} 
\usepackage{aaai25}  
\usepackage{times}  
\usepackage{helvet}  
\usepackage{courier}  
\usepackage[hyphens]{url}  
\usepackage{graphicx} 
\usepackage{natbib}  
\usepackage{caption} 
\usepackage{algorithm}
\usepackage{algpseudocode}

\algrenewcommand\algorithmicrequire{\textbf{Input:}}
\algrenewcommand\algorithmicensure{\textbf{Output:}}

\usepackage{newfloat}
\usepackage{listings}
\DeclareCaptionStyle{ruled}{labelfont=normalfont,labelsep=colon,strut=off} 
\floatstyle{ruled}
\newfloat{listing}{tb}{lst}{}
\floatname{listing}{Listing}
\usepackage{amsmath}
\usepackage{booktabs}
\usepackage{multirow}
\usepackage{colortbl}
\usepackage{amsfonts}
\usepackage{bm}

\usepackage{xcolor}

\newcommand{\answerTODO}[1][]{\textcolor{red}{\bf [TODO]}}
\newcommand{\justificationTODO}[1][]{\textcolor{red}{\bf [TODO]}}

\def\myname{SongGLM}

\title{\myname{}: Lyric-to-Melody Generation with 2D Alignment Encoding and Multi-Task Pre-Training}

\author {
    Jiaxing Yu\textsuperscript{\rm 1},
    Xinda Wu\textsuperscript{\rm 1},
    Yunfei Xu\textsuperscript{\rm 2},
    Tieyao Zhang\textsuperscript{\rm 1},
    Songruoyao Wu\textsuperscript{\rm 1},
    Le Ma\textsuperscript{\rm 1},
    Kejun Zhang\textsuperscript{\rm 1,\rm 3}\thanks{Corresponding author.}
}
\affiliations {
    \textsuperscript{\rm 1}College of Computer Science and Technology, Zhejiang University\\
    \textsuperscript{\rm 2}AI Center, Guangdong OPPO Mobile Telecommunications Corp., Ltd.\\
    \textsuperscript{\rm 3}Innovation Center of Yangtze River Delta, Zhejiang University\\
    \{yujx, wuxinda\}@zju.edu.cn, xuyunfei@oppo.com, \{kreutzer0421, wsry, maller, zhangkejun\}@zju.edu.cn
}

\begin{document}

\maketitle

\begin{abstract}
Lyric-to-melody generation aims to automatically create melodies based on given lyrics, requiring the capture of complex and subtle correlations between them. However, previous works usually suffer from two main challenges: 1) lyric-melody alignment modeling, which is often simplified to one-syllable/word-to-one-note alignment, while others have the problem of low alignment accuracy; 2) lyric-melody harmony modeling, which usually relies heavily on intermediates or strict rules, limiting model's capabilities and generative diversity. In this paper, we propose \myname{}, a lyric-to-melody generation system that leverages 2D alignment encoding and multi-task pre-training based on the General Language Model (GLM) to guarantee the alignment and harmony between lyrics and melodies. Specifically, 1) we introduce a unified symbolic song representation for lyrics and melodies with word-level and phrase-level (2D) alignment encoding to capture the lyric-melody alignment; 2) we design a multi-task pre-training framework with hierarchical blank infilling objectives (n-gram, phrase, and long span), and incorporate lyric-melody relationships into the extraction of harmonized n-grams to ensure the lyric-melody harmony. We also construct a large-scale lyric-melody paired dataset comprising over 200,000 English song pieces for pre-training and fine-tuning. The objective and subjective results indicate that \myname{} can generate melodies from lyrics with significant improvements in both alignment and harmony, outperforming all the previous baseline methods. 
\end{abstract}

\section{Introduction}

Lyric-to-melody generation, which aims to automatically generate melodies from given lyrics, has attracted lots of attention from both academia and industry. When creating melodies, capturing the complex and subtle lyric-melody correlations is crucial. Previous works~\cite{watanabe2018melody,bao2019neural,yu2021conditional,ju2021telemelody,sheng2021songmass,lv2022re,zhang2022relyme, ding2024songcomposer,wang2024mudit,yu2024suno} in this field have achieved great progress in capturing these correlations, but still encounter two primary challenges: lyric-melody alignment modeling and harmony modeling.

\begin{figure}[t]
    \centering
    \includegraphics[width=0.47\textwidth]{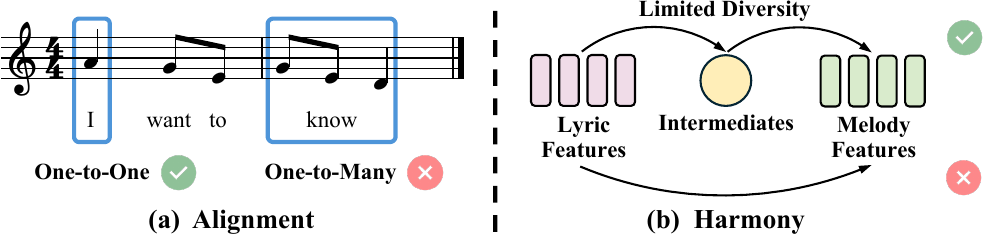}
    \caption{Illustration of lyric-to-melody generation challenges: alignment modeling and harmony modeling.}
    \label{fig:challenges}
\end{figure}

1) Lyric-melody alignment modeling. Lyric-melody alignment denotes the quantitative relationships between syllables/words and notes, particularly the number of notes mapped to a single syllable or word, which has a significant impact on the richness and singability of a song.
Since most existing works~\cite{yu2021conditional,ju2021telemelody,lv2022re,zhang2022relyme} only explore the one-syllable/word-to-one-note (one-to-one) alignment, few works consider the one-syllable/word-to-multiple-notes (one-to-multiple) alignment. 
\citet{bao2019neural} predicted the number of notes corresponding to the given syllable with a greedy alignment strategy.
\citet{sheng2021songmass} utilized sentence-level and token-level attention masks to achieve alignment between word/syllable and note.
However, these methods usually suffer from low alignment accuracy, attributable to their indirect ways of learning the lyric-melody alignment.
Consequently, it is essential to introduce a unified representation for lyrics and melodies that directly captures the lyric-melody alignment. 

2) Lyric-melody harmony modeling. Lyric-melody harmony refers to the qualitative relationships between syllables/words and notes, emphasizing their feature coherence, which is crucial for the rhythmic and structural consistency of a song.
\citet{ju2021telemelody} proposed templates which consist of tonality, chord, rhythm, and cadence, serving as a bridge between lyrics and melodies to improve the harmony. 
\citet{lv2022re} extracted key features from lyrics, including tonality, rhythm, chord, and structure, and leveraged these features as the query to retrieve and concatenate pre-generated melody segments. 
\citet{zhang2022relyme} introduced several rules about lyric-melody relationships from the perspectives of tone, rhythm, and structure, and integrated them into the decoding step of lyric-to-melody generation models.
These approaches rely heavily on either intermediates (templates or keys) or strict rules which limit model’s capabilities and generative diversity.
Therefore, proposing a method that ensures the lyric-melody harmony while maintaining generative creativity is essential.

Specifically, to address the challenge of lyric-melody alignment modeling, we introduce a unified symbolic song representation that provides a comprehensive method to encode lyric and melodic information. For both words in lyrics and notes in melodies, the representation consists of three types of attributes: generic, content-related, and alignment-related. Generic attributes refer to basic properties that apply across both words and notes, such as token types. Content-related attributes contain distinct elements describing words or notes, such as the textual contents of words and the musical features of notes. 
Alignment-related attributes include word and phrase level alignment ids (word ID and phrase ID), serving as 2D alignment encoding that directly provide hierarchical alignment information between lyrics and melodies. 

To handle the challenge of lyric-melody harmony modeling, we propose a multi-task pre-training framework based on GLM~\cite{du2021glm} that enables the model to capture multi-scale, multi-dimensional harmony between lyrics and melodies.
We concatenate the word sequence from the lyrics with the note sequence from the melody, using the word sequence as a condition.
Then, we create hierarchical blank infilling objectives (n-gram, phrase, and long span) from the perspective of word, phrase and song, to jointly pre-train the model by blanking out continuous spans of tokens from the note sequence and contextually reconstructing these spans.  
Since different n-grams contribute differently to the lyric-melody harmony, we explore the interaction of lyric and melodic features, including syllable stress, melodic peak and rhythm skeleton, and introduce two lyric-melody relationships between them. By incorporating these relationships into the process of n-gram extraction, we can select harmonized n-grams that best represent the significant and repeating patterns in the lyric-melody harmony.
Furthermore, we construct a large-scale lyric-melody paired dataset based on MelodyNet~\cite{wu2023melodyglm}, that contains more than 200,000 English song pieces for pre-training and fine-tuning. 

Our contributions are summarized as follows: 1) We propose \myname{}, a lyric-to-melody generation system that effectively tackles the challenges of lyric-melody alignment and harmony modeling. 2) We design 2D alignment encoding through a unified symbolic song representation to ensure the alignment between generated melodies and corresponding lyrics. 3) We propose a multi-task pre-training framework based on GLM with hierarchical blank infilling objectives (n-gram, phrase, and long span), and incorporate lyric-melody relationships into the extraction of harmonized n-grams to improve the harmony between lyrics and melodies. 4) Objective and subjective evaluation results show that \myname{} can generate high-quality melodies from lyrics with significant improvements in both alignment and harmony, outperforming all the previous baseline methods. This highlights the effectiveness of 2D alignment encoding and multi-task pre-training in lyric-to-melody generation.

\section{Background}

\subsection{Lyric-to-Melody Generation}

Over the past few years, there have been advancements in deep learning approaches for lyric-to-melody generation.
\citet{bao2019neural}, \citet{lee2019icomposer}, and \citet{yu2021conditional} adopted end-to-end models to generate melodies from lyrics. 
These methods cannot fully capture the relationships between lyrics and melodies due to the limited availability of paired lyric-melody dataset.
To address this issue, \citet{ju2021telemelody} divided the generation process into two stages, lyric-to-template and template-to-melody, to leverage unpaired data.
\citet{lv2022re} proposed a generation-retrieval pipeline by sharing same key features between lyrics and melodies. 
\citet{sheng2021songmass} trained lyrics generation and melody generation models separately with unpaired data and performed attention based alignment modeling.
\citet{zhang2022relyme} developed an expert system on \citet{sheng2021songmass} and \citet{ju2021telemelody} that incorporated lyric-melody relationships from the music theory to improve lyric-melody harmony.
However, the aforementioned studies are inadequate in effectively handling the complex and subtle correlations between lyrics and melodies, particularly failing to address both alignment modeling and harmony modeling concurrently.
In this paper, we propose \myname{}, a novel lyric-to-melody generation system with 2D alignment encoding and multi-task pre-training to tackle these challenges.

\subsection{Pre-Training Frameworks}

Pre-training frameworks have made significant contributions to the development of automatic music composition.
Early encoder-only frameworks, like BERT \cite{devlin2018bert}, adopted multi-layer bidirectional Transformer encoders to learn deep bidirectional representations, showing strong capabilities in music understanding tasks~\cite{wang2021musebert,zeng2021musicbert}.
MuseBERT~\cite{wang2021musebert} leveraged BERT with a specific representation that merges musical attributes and relations for better music understanding.
MusicBERT~\cite{zeng2021musicbert}, pre-training BERT with OctupleMIDI encoding and a tailored bar-level masking strategy, demonstrated strong performance across four music understanding tasks.
Later decoder-only frameworks, like GPT \cite{radford2018improving}, leveraged multi-layer unidirectional Transformer decoders to capture rich context information, which are suited for music generation tasks~\cite{ens2020mmm}.
MMM~\cite{ens2020mmm} was trained based on GPT-2~\cite{radford2019gpt2} by concatenating multiple tracks into a single sequence for conditional multi-track music generation.
MuseNet~\footnote{http://openai.com/blog/musenet} was also built upon GPT-2 that can generate 4-minute music with different instruments and various styles.
Encoder-decoder frameworks, like MASS~\cite{song2019mass}, integrated encoder and decoder modules for better understanding and generating music sequences~\cite{sheng2021songmass}.
SongMASS~\cite{sheng2021songmass} utilized the MASS pre-training method and attention-based alignment modeling for automatic song writing.
Recently, more and more innovative and efficient frameworks~\cite{dong2019unified,raffel2020exploring,du2021glm,touvron2023llama} have been proposed and adopted in different downstream tasks.
Among them, GLM~\cite{du2021glm}, based on a unified encoder-decoder framework with autoregressive blank infilling, has shown promising results in music generation tasks~\cite{wu2023melodyglm}. 
In this paper, we exploit the powerful capabilities of GLM, and build \myname{} to address the challenges encountered in lyric-to-melody generation.

\section{Method}

\subsection{System Overview}

\begin{figure}[h]
    \centering
    \includegraphics[width=0.475\textwidth]{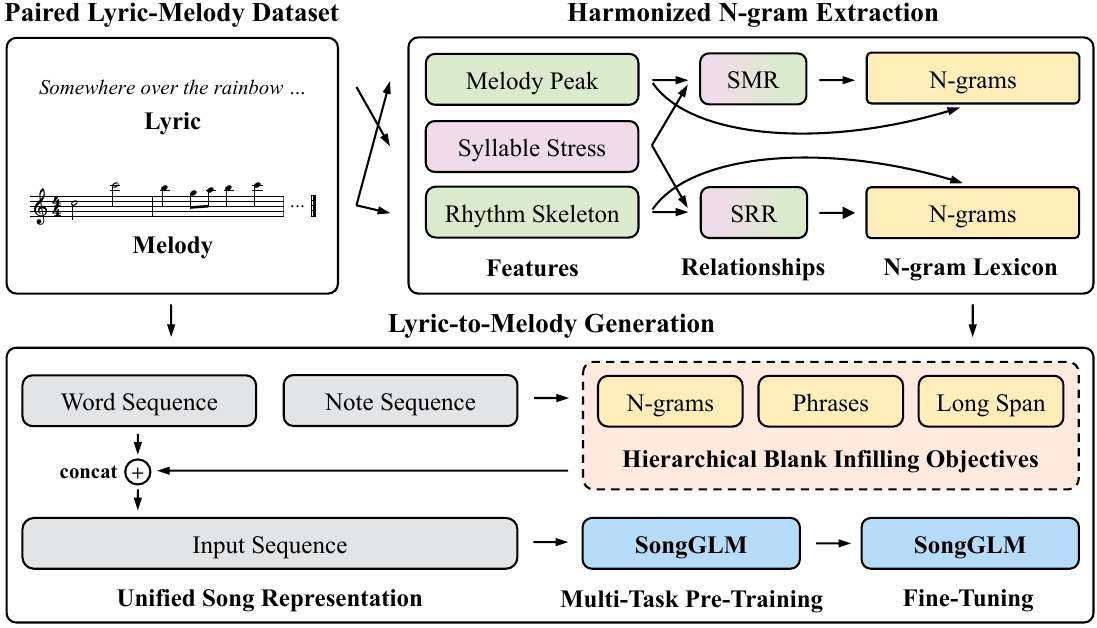}
    \caption{An overview of \myname{}. It includes two stages: harmonized n-gram extraction (detailed in Section~\ref{subsec:harmonized_n-gram_extraction}) and lyric-to-melody generation (detailed in Section~\ref{subsec:lyric-to-melody_generation}).}
    \label{fig:overview}
\end{figure}

An overview of \myname{} is shown in Figure~\ref{fig:overview}. Given the paired lyric-melody dataset, we first establish two relationships between lyrics and melodies based on their representative features, and incorporate these relationships into n-gram extraction to select the most harmonized n-grams. Then, we introduce a unified symbolic song representation with 2D alignment encoding and adopt a multi-task pre-training framework that employs hierarchical blank infilling objectives for lyric-to-melody generation. 
In the following subsections, we describe the details of harmonized n-gram extraction and lyric-to-melody generation.

\subsection{Harmonized N-gram Extraction}
\label{subsec:harmonized_n-gram_extraction}

N-grams are widely used in Music Information Retrieval for understanding and generating tasks~\cite{zheng2017music,xiao2020ernie,wu2023melodyglm}. 
However, existing n-gram extraction methods~\cite{zeng2021musicbert,wu2023melodyglm} often struggle to effectively capture the harmony between lyrics and melodies when applied to lyric-to-melody generation tasks.
To address this challenge, we first introduce the three most representative features of lyrics and melodies, and establish the qualitative relationships between them based on these features. Then, we propose a novel n-gram extraction strategy that incorporates these relationships to select the most harmonized n-grams.

\subsubsection{Features}
\label{sub:features}
\
\newline
We describe three key features of lyrics and melodies and the corresponding extraction functions ($f_w$ or $f_n$) across the dimension of word, pitch and time, specifically: \textbf{syllable stress}, \textbf{melodic peak}, and \textbf{rhythm skeleton}. 

\begin{table}[htbp]
    \small
	\centering
        \caption{Examples of syllable stress within words.}
	\label{tab:syllable_stress} 
	\begin{tabular}{p{1.45cm} p{3cm} p{2.5cm}}
            \toprule
            Words & Phoneme & Syllable Stress \\
            \midrule
            "have" & HH \  AE \  V & AE - Primary Stress \\
            "apple" & AE \  P \  AH \  L & AE - Primary Stress \par AH - Unstressed \\
            "banana" & B \  AH \  N \  AE \  N \  AH & AH - Unstressed \par AE - Primary Stress \par AH - Unstressed \\
            "watermelon" & W \  AO \  T \  ER \  M \  EH \  L \  AH \  N & AO - Primary Stress \par ER - Unstressed \par EH - Secondary Stress \par AH - Unstressed \\
            \bottomrule
        \end{tabular}
\end{table}

\noindent{\textbf{Syllable Stress. }}
Syllable stress~\cite{ladefoged1958syllables} refers to the emphasis placed on particular syllables within words. The sequence of syllable stress represents the rhythmic pattern of lyrics, and plays a crucial role in lyric-to-melody generation. According to the CMU Pronouncing Dictionary~\footnote{https://speech.cs.cmu.edu/cgi-bin/cmudict}, each syllable stress can be categorized into three levels on an ordinal scale: Unstressed, Primary Stress, and Secondary Stress. Table~\ref{tab:syllable_stress} presents several examples of syllable stress within words from the CMU Pronouncing Dictionary. For the ``Syllable Stress'' feature, we define $f_w$ as follows: 
\begin{equation}
    f_w(W) = [s_1,\ldots,s_y]
\end{equation}
where $s_i \in \{0,1,2\}$ represents the stress level of the $i^{th}$ syllable in a word, with 0 indicating Unstressed, 1 indicating Primary Stress, and 2 indicating Secondary Stress.

\begin{figure}[h]
    \centering
    \includegraphics[width=0.48\textwidth]{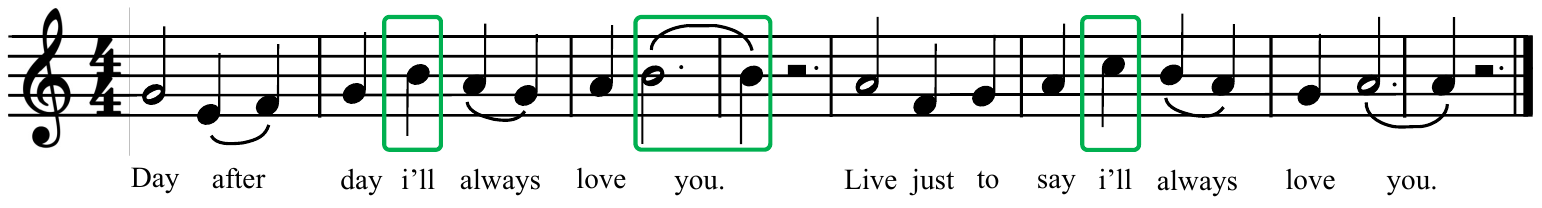}
    \caption{Examples of melodic peaks in the song.}
    \label{fig:melodic_peak}
\end{figure}

\noindent{\textbf{Melodic Peak. }}
Melodic peaks~\cite{eitan2016highpoints} refer to the notes with a higher pitch compared to the preceding and subsequent notes. The sequence of melodic peaks describes the movement pattern of the melody among high pitches. Figure~\ref{fig:melodic_peak} illustrates examples of melodic peaks (green blocks) in the song. For the ``Melodic Peak'' feature, given the pitch sequence $P = [p_1,\ldots,p_n]$ of the melody, we define $f_{n.{MP}}$ for the $i^{th}$ note as follows:

\begin{equation}
    f_{n.{MP}}(N) = \begin{cases} 
    1, & 1 < i < n \text{ and } p^{i} > p^{i-1} \text{ and } p^{i} > p^{i+1} \\
    0, & \text{otherwise}
    \end{cases}
\end{equation}

\begin{figure}[h]
    \centering
    \includegraphics[width=0.479\textwidth]{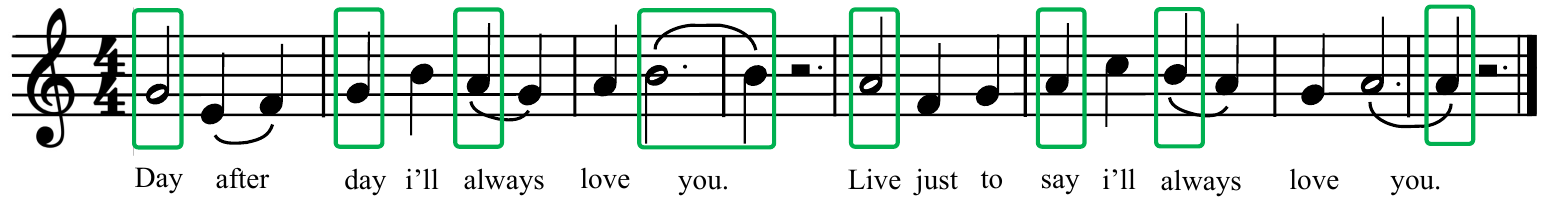}
    \caption{Examples of the rhythm skeleton in the song.}
    \label{fig:rhythm_skeleton}
\end{figure}

\noindent{\textbf{Rhythm Skeleton. }}
The rhythm skeleton~\cite{zhang2023wuyun} represents a set of specific notes that are acoustically more prominent than others, due to the joint effect of meter and rhythm on the time dimension. Following~\cite{zhang2023wuyun}, we extract metrical accents, agogic accents on metrical accents, and agogic accents on syncopations as the rhythm skeleton. Figure~\ref{fig:rhythm_skeleton} illustrates examples of rhythm skeleton (green blocks) in the song. For the ``Rhythm Skeleton'' feature, given the rhythm skeleton sequence $RS = [N_1,\ldots,N_z]$ of the melody, we define $f_{n.{RS}}$ for the $i^{th}$ note as follows:

\begin{equation}
    f_{n.{RS}}(N) = \begin{cases} 
    1, & N_i \in RS \\
    0, & N_i \notin RS
    \end{cases}
\end{equation}

\begin{figure}[h]
    \centering
    \includegraphics[width=0.48\textwidth]{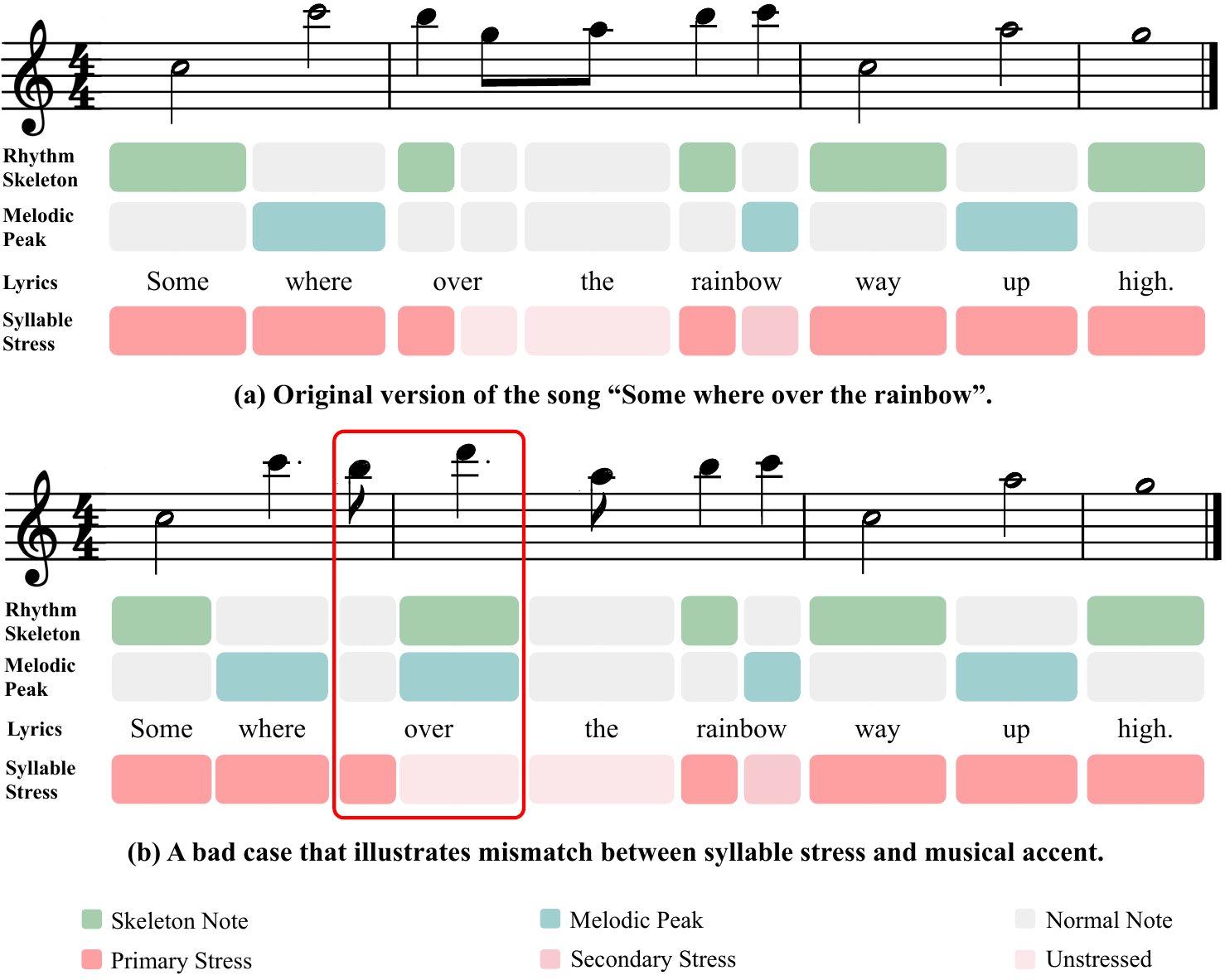}
    \caption{An example of lyric and melodic features, as well as the relationships between them.}
    \label{fig:relationships}
\end{figure}

\subsubsection{Relationships}
\
\newline
Since stressed syllables are often associated with musically accented notes~\cite{nichols2009relationships}, we reveal the mechanism of interaction between lyric and melodic features and introduce two relationships.

\noindent{\textbf{Syllable Stress and Melodic Peak (SMR). }}
Composers usually employ melodic peaks to emphasize certain syllables in songwriting. Specifically, as shown in Figure~\ref{fig:relationships}(a), a high level of syllable stress tends to occur with the melodic peak. 

\noindent{\textbf{Syllable Stress and Rhythm Skeleton (SRR). }}
Rhythm skeleton is another method by which composers highlight specific words in a song. For instance, as illustrated in Figure~\ref{fig:relationships}(a), syllables with high stress levels are often associated with the note in the rhythm skeleton.

The correspondence between syllable stress and melodic accents can significantly improve the harmony between lyrics and melodies. On the other hand, if a melodic accent mismatches with syllable stress (red block in Figure~\ref{fig:relationships}(b)), it may disrupt the natural flow of the song, potentially resulting in disharmony that can detract from the overall listening experience.

\subsubsection{Extraction Strategy}
\label{sub:extraction_strategy}
\
\newline
To capture the above relationships and ensure harmony between lyrics and melodies, we propose a novel n-gram extraction strategy. 
This strategy involves calculating a composite score for each n-gram, which includes both a melodic score and a lyric-melody relationship score. N-grams with high composite scores are selected as harmonized n-grams. The details of this strategy are outlined below.
 
Given a word sequence $W$ of lyrics, a note sequence $N$ of melody, and paired feature extraction functions $f_w$ and $f_n$ ($f_{n.{MP}}$ or $f_{n.{RS}}$) from lyric-melody relationships, we denote each word-note feature pair as a joint uni-gram $\{(f_w(w), f_n(N_w)) \mid w \in W, N_w \subseteq N\}$, where $N_w$ represents the set of notes corresponding to the single word $w$. Subsequently, we extract joint n-grams for $n$ ranging from 2 to 12, comprising lyric n-grams $f_w(W_n)$ with $n$ words and melodic n-grams $f_n(N_{W_n})$. 
Furthermore, we compute t-statistic scores $s_t$~\cite{xiao2020ernie} for the lyric and melodic n-grams separately, $s_l$ and $s_m$. Each melodic n-gram $f_n(N_{W_n})$ is associated with a set of $m$ distinct lyric n-grams $F_w(W_n^m)=\{f_w(W_n^1),\ldots,f_w(W_n^m)\}$ from different joint n-grams. 
Finally, the score $s$ of a joint n-gram $\{(f_w(W_n), f_n(N_{W_n})) \mid W_n \subseteq W, N_{W_n} \subseteq N\}$ consists of two parts (the melodic n-gram t-statistic score $s_m$ and the lyric-melody relationship score $s_{lm}$), which is defined as:
\begin{equation}
    \begin{aligned}
        s=s_m + s_{lm}
    \end{aligned}
\end{equation}

\begin{equation}
    \begin{aligned}
        s_l=s_t(f_w(W_n))
    \end{aligned}
\end{equation}

\begin{equation}
    \begin{aligned}
        s_m=s_t(f_n(N_{W_n}))
    \end{aligned}
\end{equation}

\begin{equation}
    \begin{aligned}
        s_{lm}=C(F_w(W_n^m)) \cdot \frac{1}{m} \sum_{i=1}^{m} s_l^i
    \end{aligned}
\end{equation}

\begin{equation}
    \begin{aligned}
        C(F_w(W_n^m)) = 
        \begin{cases}
        1, & m = 1 \\
        1 - H'(F_w(W_n^m)), & m > 1 \\
        \end{cases}
    \end{aligned}
\end{equation}

\begin{equation}
    \begin{aligned}
        H'(F_w(W_n^m)) = \frac{-\sum_{i=1}^{m} p(W_n^i) \log p(W_n^i)}{\log m}
    \end{aligned}
\end{equation}
where $p$ represents the occurrence probability of a given lyric n-gram among all corresponded lyric n-grams to the melodic n-gram, $C$ represents the concentration of the lyric n-gram set associated with the melodic n-gram, derived from the normalized entropy $H'$. The higher the concentration $C$, the better the joint n-gram represents a significant and repeating pattern in the lyric-melody relationship, thereby more effectively influencing the harmony between lyrics and melodies. 
Based on their scores, we select the top 25\% of joint n-grams as harmonized n-grams to construct the final n-gram lexicon for word-level sampling in lyric-to-melody generation.

\begin{table*}[h]
    \small
	\centering
        \begin{tabular*}{\textwidth}{p{2.2cm} p{2cm} p{10.5cm} l}
            \toprule
            Attribute Type & Attribute Name & Representation & \# Size \\
            \midrule
            \multirow{6}{*}{Content-Related} & Bar & Bar\_Val (Val $ \in \{x \in \mathbb{N} \mid 0 \leq x \leq 127\}$) & 128 \\
            & Position & Pos\_Val (Val $ \in \{0, 30, 60, \ldots, 1890\} \cup \{0, 40, 80, \ldots, 1880\}$) & 96 \\
            & Pitch & Pitch\_Val (Val $\in \{x \in \mathbb{N} \mid 0 \leq x \leq 127\}$) & 128 \\
            & Duration & Dur\_Val (Val $\in \{30, 60, 90, \ldots, 1920\} \cup \{40, 80, 160, 320, 640\}$) & 69 \\
            & Tempo & Large (Tempo $< 60$), Larghetto (Tempo $\in [60, 66)$), Adagio (Tempo $\in [66, 76)$), \par Andante (Tempo $\in [76, 108)$), Moderato (Tempo $\in [108, 120)$), \par Allegro (Tempo $\in [120, 168)$), Presto (Tempo $\geq 168$) & 7 \\
            & Text & "the", "i", "a", "you", "and", "to", $\ldots$ , "ratio" & 23,648 \\
            \midrule
            \multirow{2}{*}{Alignment-Related} & Word ID & Word\_Val (Val $ \in \{x \in \mathbb{N} \mid 0 \leq x \leq 255\}$) & 256 \\
            & Phrase ID & Phrase\_Val (Val $ \in \{x \in \mathbb{N} \mid 0 \leq x \leq 127\}$) & 128 \\
            \midrule
            Generic & Token Type & Word, Note & 2 \\
            \bottomrule
        \end{tabular*}
        \caption{Details of attributes in our unified symbolic song representation.}
	\label{tab:attributes}
\end{table*}

\subsection{Lyric-to-Melody Generation}
\label{subsec:lyric-to-melody_generation}

On top of the above extracted harmonized n-grams, we build \myname{} upon GLM~\cite{du2021glm} with a single Transformer-based encoder-decoder framework for lyric-to-melody generation, as shown in Figure~\ref{fig:detailed_framework}. 
In the pre-training stage, we adopt a multi-task pre-training framework with hierarchical blank infilling objectives. In the fine-tuning and inference stage, we utilize causal language modeling to predict the next note sequentially from left to right.

\begin{algorithm}
    \caption{Musical Phrase Boundary Recognition}
    \label{alg:musical_phrase_boundary_recognition}
    \begin{algorithmic}[1]
        \Require
            \Statex Word sequence of lyrics $W = \{W_1,\ldots,W_m\}$
            \Statex Note sequence of melody $N = \{N_1,\ldots,N_n\}$
        \Ensure
            \Statex Musical phrase ending notes $E = \{E_1,\ldots,E_x\}$

        \Function{LyricsBasedRecognition}{$W$}
            \State $LE \gets \{\}$
            \For{$i \gets 0$ to $m$}
                \If{$W_i$ contains punctuation marks}
                    \State $LE$.append($W_i$)
                \EndIf
            \EndFor
            \State \Return $LE$
        \EndFunction

        \Function{MelodyBasedRecognition}{$N$}
            \State $L \gets  LongNotes(N)$
            \State $R \gets RestNotes(N)$
            \State $ME \gets L \cup R$
            \State $i \gets 1$
            
            \While{$i < n$}
                \If{$ME_{i-1}$ and $ME_{i}$ are adjacent in $N$}
                    \State $D \gets  |ME_{i-1}.duration - ME_{i}.duration|$
                    \If{$D > 240$}
                        \State remove $ME_{i}$
                    \Else
                        \State remove $ME_{i-1}$
                    \EndIf
                \EndIf
                \State $i \gets i + 1$
            \EndWhile
            \State \Return $ME$
        \EndFunction

        \State $LR \gets \Call{LyricsBasedRecognition}{W}$
        \State $MR \gets \Call{MelodyBasedRecognition}{N}$
        
        \State $PR \gets \frac{len(LR)}{len(W)}$ \Comment{calculate the punctuation mark ratio}
        \If{$PR < 0.1$}
            \State \Return $MR$
        \Else
            \State \Return $LR$
        \EndIf
    \end{algorithmic}
\end{algorithm}

\subsubsection{Unified Symbolic Song Representation}
\label{subsec:unified_symbolic_song_representation}
\
\newline
Inspired by OctupleMIDI~\cite{zeng2021musicbert}, We design a unified symbolic song representation for lyric-to-melody generation that allows the model to learn the lyric-melody alignment in an efficient and direct way. It consists of three different types of tokens: \textit{\textbf{Word}}, \textit{\textbf{Note}} and \textit{\textbf{Special}}, each containing three sets of attributes: content-related, alignment-related, and generic. We list all the attributes in Table~\ref{tab:attributes}. For each token, we consolidate the attributes into a single compound token to reduce the sequence length.

\begin{figure*}[h]
    \centering
    \includegraphics[width=\textwidth]{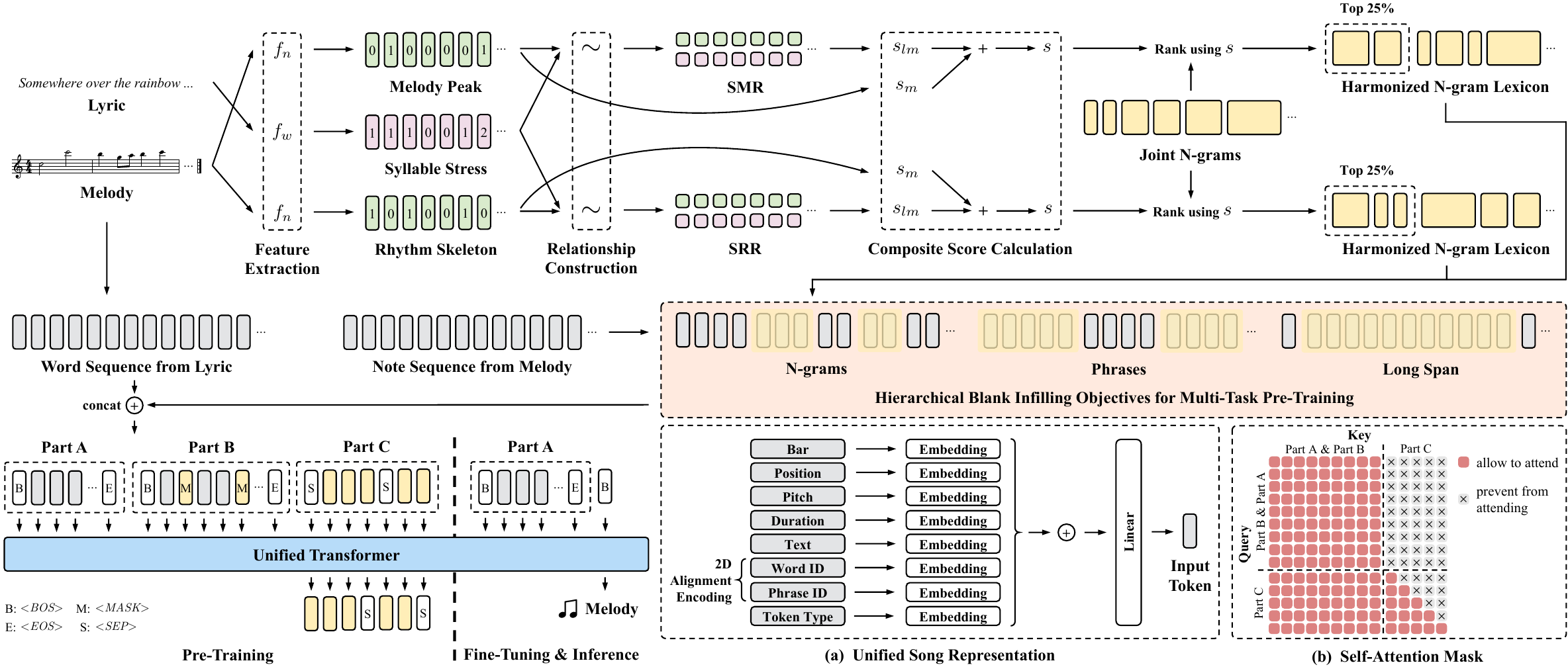}
    \caption{Illustration of our proposed \myname{}. In the harmonized n-gram extraction stage (top), we construct two lyric-melody relationships based on their features, and incorporate these relationships into n-gram extraction to select the top 25\% as harmonized n-grams. In the lyric-to-melody generation stage (bottom), we leverage the extracted n-grams and propose three hierarchical blank infilling objectives for multi-task pre-training. We adopt casual language modeling for fine-tuning and inference. (a) Unified song representation for both lyrics and melodies. (b) Self-attention mask in the unified transformer.}
    \label{fig:detailed_framework}
\end{figure*}

For \textit{\textbf{Word}} and \textit{\textbf{Note}} tokens, we assign the same alignment-related and generic attributes but different content-related attributes. 
Specifically, alignment-related attributes include two alignment ids, as shown in Figure~\ref{fig:detailed_framework}(a). The first alignment id represents word-level alignment, called Word ID. For each \textit{\textbf{Word}} token, it denotes the position in the word sequence, starting from 0 to the total length - 1. For each \textbf{\textit{Note}} token, it equals to the Word ID of the word corresponding to the note. The second alignment id represents phrase-level alignment, named Phrase ID. For both tokens, the Phrase IDs refer to the musical phrase to which they belong. 
The above two alignment ids are encoded into embedding vectors, serving as \textbf{2D alignment encoding} to guarantee the hierarchical alignments between words and notes. Generic attributes include token types, which enhance the model's capacity to differentiate between \textit{\textbf{Word}} and \textit{\textbf{Note}}. And for content-related attributes, \textit{\textbf{Note}} tokens comprise five musical elements: bar, position, pitch, duration, and tempo, while \textit{\textbf{Word}} tokens contain the text of the word.
We only select words that are included in the CMU Pronouncing Dictionary and sorted them according to their frequency of occurrence in the lyrics. To facilitate computational modeling, we set content-related attributes of the \textit{\textbf{Word}} token to \textit{None} in the \textit{\textbf{Note}} token, and vice versa.

For \textit{\textbf{Special}} tokens, we adopt five special delimiter symbols: \textit{\textless BOS\textgreater}, \textit{\textless EOS\textgreater}, \textit{\textless MASK\textgreater}, \textit{\textless PAD\textgreater}, and \textit{\textless SEP\textgreater}. Similar to the \textit{\textbf{Word}} and \textit{\textbf{Note}} tokens, every \textit{\textbf{Special}} token contains all attributes, each bearing the same value as itself.

\subsubsection{Multi-Task Pre-Training}
\label{subsec:multi-task_pre-training}
\
\newline
Multi-task pre-training has been shown to enhance model's performance in a variety of tasks~\cite{sun2021ernie3,wu2023melodyglm}. Meanwhile, autoregressive blank infilling is an effective pre-training approach for language models~\cite{du2021glm,wu2023melodyglm}. Following their success, we implement a multi-task pre-training framework with hierarchical autoregressive blank infilling objectives in \myname{}.

Autoregressive blank infilling involves blanking out continuous spans of tokens from the input sequence and contextually reconstructing these spans during model training. 
Given an input sequence $S = [W_1,\ldots,W_m, N_1,\ldots,N_n]$, multiple token spans $s = \{s_1,\ldots,s_k\}$ are sampled from the note sequence $N$. Each span $s_i$ corresponds to a series of consecutive tokens $[s_{i,1},\ldots,s_{i,l_i}]$ in $N$, and is replaced with a single special token \textit{\textless MASK\textgreater}, forming a corrupted token sequence $S_{corrupt}$. The model is trained to predict the missing tokens within the spans from the corrupted token sequence in an autoregressive way, with access to the corrupted token sequence and previously predicted spans. 
Formally, the generation probability of the $i^{th}$ masked span is defined as:
\begin{equation}
    \begin{aligned}
        p_{\theta}(s_i|S_{corrupt},s_{<i}) = \prod_{j=1}^{l_i}p_\theta(s_{i,j}|S_{corrupt},s_{<i},s_{i,<j})
    \end{aligned}
\end{equation}
And an autoregressive blank infilling objective is performed by minimizing the negative likelihood (loss) as follows: 
\begin{equation}
    \begin{aligned}
        -{\log} p_{\theta}(s|S_{corrupt}) = -\sum_{s_i{\in}s}\sum_{s_{i,j}{\in}s_i}{\log} p_{\theta}(s_{i,j}|S_{corrupt})
    \end{aligned}
\end{equation}

We construct three hierarchical autoregressive blank infilling objectives for pre-training to capture the multi-scale, multi-dimensional harmony between lyrics and melodies.

\noindent{\textbf{Word-Level. }}
Based on the extracted n-gram lexicon, we randomly sample two types of harmonized n-grams from the note sequence with the Maximum Matching Algorithm \cite{xiao2020ernie}. The total length of sampled n-grams constitutes 15\% of the note sequence. We replace each sampled n-gram with 1) the \textit{\textless MASK\textgreater} token 80\% of the time, 2) a random n-gram 10\% of the time, and 3) the original n-gram 10\% of the time. These objectives aims to capture word-note level harmony between lyrics and melodies.

\noindent{\textbf{Phrase-Level. }}
Multiple musical phrases are sampled from the note sequence, with the total length accounting for 50\% of the original note sequence length. We consider both lyric and melodic information for musical phrase boundary recognition. The detailed detection algorithm
is shown in Algorithm~\ref{alg:musical_phrase_boundary_recognition}. This objective aims to capture lyric-phrase level harmony between lyrics and melodies, as well as to ensure the coherence of melodic contexts.
    
\noindent{\textbf{Song-Level. }} We sample a single long span that covers 50\% of the original note tokens. This objective aims to improve the overall harmony between lyrics and melodies, and enhance the model's ability of melodic structure modeling.

The loss of our proposed multi-task pre-training objectives is defined as:
\begin{equation}
    \begin{aligned}
        \mathcal{L}=\mathcal{L}_{SMR}+\mathcal{L}_{SRR}+\mathcal{L}_{Phrase}+\mathcal{L}_{Song}
    \end{aligned}
\end{equation}

\subsubsection{Sequence Modeling}
\label{subsec:sequence_modeling}
\
\newline
\noindent{\textbf{Pre-Training. }}
In the pre-training stage, the input sequence $S$ contains three parts: Part A is the word sequence $W$, Part B is the corrupted note sequence, and Part C consists of the masked spans with each separated by a \textit{\textless SEP\textgreater} token. 
Tokens in Part A \& Part B form the corrupted sequence $S_{corrupt}$, and can attend to each other. Part C tokens can only attend to preceding tokens, and tokens in Part A \& Part B. 
Figure~\ref{fig:detailed_framework}(b) illustrates how attention weight is modified through the self-attention mask to control the token's attention.
Formally, $W_A$ and $M$ are described as:
\begin{equation}
    \begin{aligned}
        W_A = softmax(\frac{QK^T}{\sqrt{d_k}} + M)
    \end{aligned}
\end{equation}
\begin{equation}
    \begin{aligned}
        M_{ij} = \begin{cases} 
        0, & \text{allow to attend}  \\
        -\infty, & \text{  prevent from attending } \\
        \end{cases}
    \end{aligned}
\end{equation}
With this mechanism, our unified model effectively learns a bidirectional encoder for Part A \& Part B, and a unidirectional decoder for Part C.

\noindent{\textbf{Fine-Tuning and Inference. }}
In the fine-tuning and inference stage, we employ causal language modeling. The input sequence begins with the word sequence (Part A) and a \textit{\textless BOS\textgreater} token (indicating the start of the note sequence), and the model predicts the next token in an autoregressive manner until it generates an \textit{\textless EOS\textgreater} token.

\section{Experiments}

\subsection{Lyric-Melody Dataset}

A large-scale paired dataset is critical for lyric-to-melody generation models to capture lyric-melody correlations and attain superior performance. However, the current largest paired dataset~\cite{yu2021conditional} only contains 12,197 MIDI songs and lacks one-to-multiple alignment. 
In this paper, we acquire approximately 1.6 million raw MIDI data from MelodyNet \cite{wu2023melodyglm}, and construct a large-scale lyric-melody paired dataset with varied word-note alignments, including both one-to-one and one-to-multiple alignments.

\subsubsection{Data Processing}
\
\newline
To obtain high-quality MIDI songs from raw MIDI data, we perform data processing in four phases: lyric processing phase, melody processing phase, lyric-melody combined processing phase, and de-duplication phase.

\noindent{\textbf{Lyric Processing Phase. }}
First, we clean the lyrics by retaining only English letters and punctuation marks~\footnote{The punctuation marks refer to quotes, commas, colons, semicolons, periods, question marks, and exclamation marks.}, and converting the text to lowercase. Second, given the mixture of words and syllables in the lyrics,  we combine syllables into words, and then remove words not in the CMU Pronouncing Dictionary. Finally, we filter the lyrics through heuristic rules, eliminating those with a high repetition of words (greater than 20\%) and a high proportion of long/short words (greater than 50\%).

\noindent{\textbf{Melody Processing Phase. }}
First, we extract melodies with a 4/4 time signature and a constant tempo from the raw MIDI data. Second, to fit the range of human vocals, we perform octave transposition by adjusting the pitch range from C3 to C5 and quantize notes to the closest quantization grids including 16th, 32nd, 64th notes, and triplets. Finally, we filter out the empty bars in each melody and only consider melodies with at least 8 bars. 

\noindent{\textbf{Lyric-Melody Combined Processing Phase. }}
For every piece of song, we align the start time of each word to the nearest note. If multiple words correspond to the same note, only one of the words is retained. After this phase, we can obtain a lyric-melody paired dataset with word-note level alignments.

\noindent{\textbf{De-Duplication Phase. }}
We apply both internal and external de-duplication to the processed dataset by the hash value of note sequences and lyric sequences.

After data processing, the final dataset contains 206,884 English MIDI songs with 4,921.79 hours of melodies in total, which can be directly used for lyric-to-melody generation tasks. Detailed dataset statistics are shown in Table~\ref{tab:dataset_statistics}.

\begin{table}[htbp]
    \small
	\centering
        \caption{Statistics of lyric-melody dataset.}
	\label{tab:dataset_statistics} 
	\begin{tabular}{p{6.5cm} p{1.25cm}}
            \toprule
            Items & \# Size \\
            \midrule
            \# of songs & 206,884 \\
            Total duration (hours) & 4,921.79 \\
            Average duration of per song (seconds) & 85.64 \\
            Average \# of words per song & 114.56 \\
            Average \# of notes per song & 143.65 \\ 
            \bottomrule
        \end{tabular}
\end{table}

\begin{table}[htbp]
    \small
	\centering
        \caption{The model configurations of \myname{}.}
	\label{tab:model_configurations} 
	\begin{tabular}{p{3.7cm} p{1.85cm} p{1.85cm}}
            \toprule
            Configurations & \myname{}$_{\text{small}}$ & \myname{}$_{\text{base}}$ \\
            \midrule
            \# of encoder/decoder layers & 4 & 8 \\
            \# of attention heads & 8 & 8 \\
            Hidden size & 1024 & 1536 \\
            FFN inner hidden size & 2048 & 3072 \\
            Embedding sizes & & \\
            \ \ $\cdot$ \ Bar & 16 & 32 \\
            \ \ $\cdot$ \ Position & 256 & 512 \\
            \ \ $\cdot$ \ Pitch & 256 & 512 \\
            \ \ $\cdot$ \ Duration & 256 & 512 \\
            \ \ $\cdot$ \ Tempo & 128 & 256 \\
            \ \ $\cdot$ \ Text & 768 & 1024 \\
            \ \ $\cdot$ \ Word ID & 256 & 512 \\
            \ \ $\cdot$ \ Phrase ID & 256 & 512 \\
            \ \ $\cdot$ \ Token Type & 16 & 32 \\
            \midrule
            \# of trainable parameters & 57M & 200M \\
            \bottomrule
        \end{tabular}
\end{table}

\subsubsection{Data Analysis}
\begin{figure}[h]
    \centering
    \includegraphics[width=0.48\textwidth]{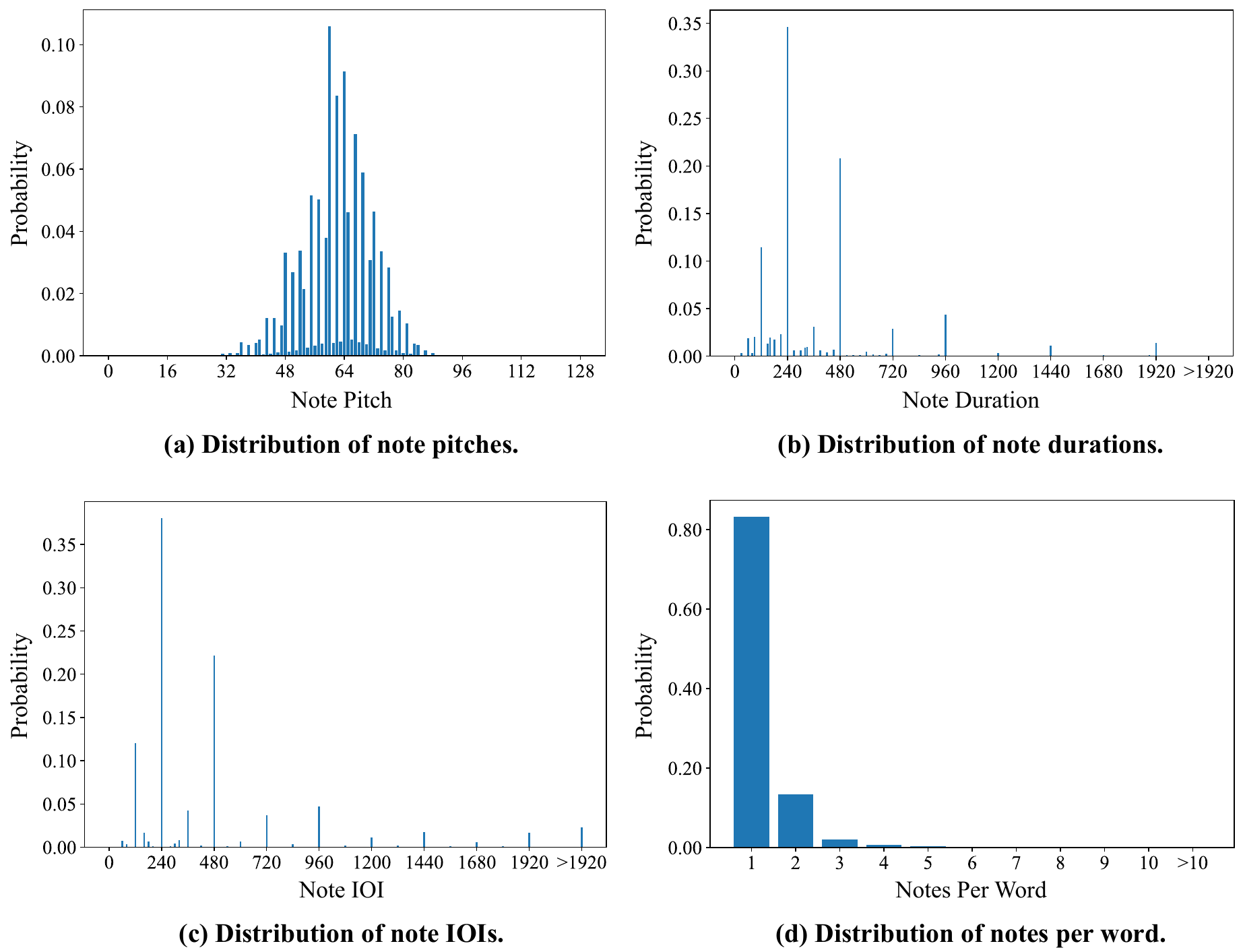}
    \caption{Distribution of music attributes in our lyric-melody dataset.}
    \label{fig:distribution}
\end{figure}

\
\newline
To analyze the musical attributes and word-note alignments within our dataset, we calculate the distribution of pitch, duration, Inter Onset Interval (IOI)~\cite{yang2020evaluation} and notes per word, as individually illustrated in Figure~\ref{fig:distribution}. It indicates that the majority of note pitches fall within the range of 48 to 72 due to octave transposition, and the presence of triplets enhances the diversity of note durations and IOIs, thus providing rich knowledge for the model to understand lyric-melody harmony. Furthermore, approximately 20\% of the words correspond to multiple notes, which is crucial for the model to capture complex alignments between lyrics and melodies.

\subsection{Model Configuration}

\myname{} uses a single Transformer \cite{vaswani2017attention} as the basic model structure, and is pre-trained in two versions: 1) \myname{}$_{\text{small}}$ on the small-scale dataset, containing 40,000 songs randomly selected from the full pre-training dataset, which aims to compare with baselines that are also pre-trained on the small-scale dataset; 2) \myname{}$_{\text{base}}$ on the full pre-training dataset, for demonstrating the best capability of \myname{} and presenting the state-of-the-art results. Detailed configurations of two \myname{} models are shown in Table~\ref{tab:model_configurations}. We adopt our proposed multi-task pre-training framework for both \myname{}$_{\text{small}}$ and \myname{}$_{\text{base}}$.

\subsubsection{Pre-training and Fine-tuning Details}
\
\newline
\myname{} is pre-trained on a single NVIDIA A100 80GB Tensor Core GPU over a total of 250,000 steps. The batch size is set to 92, with each sequence up to 768 tokens. The dropout rate is 0.1. We employ AdamW optimizer (${\beta}_1 = 0.9$, ${\beta}_2 = 0.98$, ${\epsilon} = {10^{-6}}$) with weight decay of 0.01. We use the one-cycle learning rate policy to schedule the learning rate. Specifically, it includes a warm-up phrase for the first 50,000 steps to a maximum learning rate of 0.00035, followed by a cosine decay until it reaches 250,000 total updates.

After pre-training, we fine-tune \myname{} for the lyric-to-melody generation task. It aims to generate high-quality and correlated melody from given lyrics. We set training steps to 100,000 with a batch size of 4. The learning rate is warmed up over the first 20,000 steps to a peak value of 0.00005, and then decays with cosine annealing. For inference, we apply the temperature-controlled stochastic sampling method with top-${k}$~\cite{keskar2019ctrl} (${k} = 10$, ${temperature} = 0.9$). Other parameters remain consistent with the settings used in pre-training.

\begin{table*}[h]
    \small
	\centering
        \begin{tabular*}{\textwidth}{p{3.53cm} m{2.38cm} m{2.38cm} m{2.38cm} m{2.38cm} m{2.38cm}}
            \toprule
            \multicolumn{1}{l}{\multirow{2}{*}{\textbf{Model}}} & \multicolumn{1}{c}{\textbf{Alignment}} & \multicolumn{4}{c}{\textbf{Harmony}} \\
		  \cmidrule(lr){2-2} \cmidrule(lr){3-6}
		  & \centering $D_A(\%)$~$\uparrow$ & \centering $D_P(\%)$~$\uparrow$ & \centering $D_D(\%)$~$\uparrow$ & \centering $D_{IOI}(\%)$~$\uparrow$ & \centering $MD$~$\downarrow$ \arraybackslash \\
		\midrule
            SongMASS & \centering - & \centering 87.25 $\pm$ 1.39 & \centering 75.79 $\pm$ 1.23 & \centering 81.94 $\pm$ 2.02 & \centering 8.48 $\pm$ 0.75 \arraybackslash \\
            TeleMelody & \centering - & \centering 89.62 $\pm$ 1.12 & \centering 84.55 $\pm$ 1.52 & \centering 79.38 $\pm$ 0.88 & \centering 6.36 $\pm$ 0.89 \arraybackslash \\
            ReLyMe (in SongMASS) & \centering - & \centering 90.25 $\pm$ 0.80 & \centering 84.65 $\pm$ 0.93 & \centering 86.69 $\pm$ 1.04 & \centering 6.98 $\pm$ 0.80 \arraybackslash \\
            ReLyMe (in TeleMelody) & \centering - & \centering 92.19 $\pm$ 0.72 & \centering 87.52 $\pm$ 1.02 & \centering 84.80 $\pm$ 1.17 & \centering 5.90 $\pm$ 0.88 \arraybackslash \\
            \midrule
            \myname{}$_{\text{small}}$ & \centering 94.32 $\pm$ 0.64 & \centering 96.48 $\pm$ 0.94 & \centering 95.34 $\pm$ 0.99 & \centering 93.44 $\pm$ 0.72 & \centering 4.17 $\pm$ 0.21 \arraybackslash \\
            \myname{}$_{\text{base}}$ & \centering \textbf{96.83 $\bm{\pm}$ 0.59} & \centering \textbf{96.50 $\bm{\pm}$ 0.71} & \centering \textbf{96.48 $\bm{\pm}$ 0.97} & \centering \textbf{94.10 $\bm{\pm}$ 0.93} & \centering \textbf{3.85 $\bm{\pm}$ 0.30} \arraybackslash \\
            \midrule
            \midrule

            Scratch & \centering 84.15 $\pm$ 0.69 & \centering 88.61 $\pm$ 0.83 & \centering 86.97 $\pm$ 1.09 & \centering 82.91 $\pm$ 1.03 & \centering 6.29 $\pm$ 0.84 \arraybackslash \\
            CLM & \centering 90.20 $\pm$ 0.66 & \centering 91.90 $\pm$ 0.68 & \centering 90.69 $\pm$ 1.04 & \centering 88.09 $\pm$ 0.99 & \centering 4.95 $\pm$ 0.32 \arraybackslash \\
            \ \  -- \ 2D Alignment Encoding & \centering 83.57 $\pm$ 0.89 & \centering 92.12 $\pm$ 0.84 & \centering 87.40 $\pm$ 1.02 & \centering 84.61 $\pm$ 1.11 & \centering 5.67 $\pm$ 0.60 \arraybackslash \\
            Random & \centering 93.87 $\pm$ 0.68 & \centering 92.84 $\pm$ 0.65 & \centering 92.06 $\pm$ 1.11 & \centering 89.07 $\pm$ 1.21 & \centering 4.68 $\pm$ 0.52 \arraybackslash \\
            Harmonized N-gram & \centering 93.74 $\pm$ 0.51 & \centering 94.11 $\pm$ 0.67 & \centering 93.65 $\pm$ 1.09 & \centering 91.88 $\pm$ 1.05 & \centering 4.40 $\pm$ 0.48 \arraybackslash \\
            \ \ -- \ SMR & \centering 93.67 $\pm$ 0.60 & \centering 91.77 $\pm$ 0.78 & \centering 92.98 $\pm$ 1.06 & \centering 91.13 $\pm$ 1.03 & \centering 4.79 $\pm$ 0.41 \arraybackslash \\
            \ \ -- \ SRR & \centering 93.54 $\pm$ 0.59 & \centering 93.93 $\pm$ 0.68 & \centering 90.81 $\pm$ 1.03 & \centering 89.04 $\pm$ 1.10 & \centering 4.86 $\pm$ 0.45 \arraybackslash \\
            Phrase & \centering 92.95 $\pm$ 0.58 & \centering 94.09 $\pm$ 0.66 & \centering 93.71 $\pm$ 1.21 & \centering 91.72 $\pm$ 1.17 & \centering 4.66 $\pm$ 0.25 \arraybackslash \\
            Long & \centering 93.54 $\pm$ 0.67 & \centering 94.24 $\pm$ 0.63 & \centering 93.67 $\pm$ 1.06 & \centering 91.68 $\pm$ 1.10 & \centering 4.60 $\pm$ 0.49 \arraybackslash \\
            
            \bottomrule
        \end{tabular*}
        \caption{Objective results of \myname{} with different settings and baseline systems (Mean $\pm$ SD). SMR refers to syllable stress and melodic peak relationship, and SRR refers to syllable stress and rhythm skeleton relationship.}
        \label{tab:objective_results}
\end{table*}

\begin{table*}[h]
    \small
	\centering
        \begin{tabular*}{\textwidth}{p{3.28cm} m{1.55cm} m{1.6cm} m{1.75cm} m{1.75cm} m{1.55cm} m{1.55cm} m{1.55cm}}
            \toprule
            \multicolumn{1}{l}{\multirow{2}{*}{\textbf{Model}}} & \multicolumn{3}{c}{\textbf{Melody}} & \multicolumn{4}{c}{\textbf{Melody + Lyrics}} \\
		  \cmidrule(lr){2-4} \cmidrule(lr){5-8}
		  & \centering Richness & \centering Consistency & \centering     Listenability & \centering Rhythmicity & \centering Structure & \centering Singability & \centering Overall \arraybackslash \\
		\midrule
            SongMASS & \centering 6.10 $\pm$ 0.39 & \centering 6.03 $\pm$ 0.36 & \centering 5.91 $\pm$ 0.40 & \centering 5.76 $\pm$ 0.39 & \centering 5.78 $\pm$ 0.28 & \centering 5.68 $\pm$ 0.38 & \centering 5.77 $\pm$ 0.32 \arraybackslash \\
            TeleMelody & \centering 6.54 $\pm$ 0.36 & \centering 6.38 $\pm$ 0.35 & \centering 6.39 $\pm$ 0.42 & \centering 6.33 $\pm$ 0.42 & \centering 6.26 $\pm$ 0.38 & \centering 6.38 $\pm$ 0.42 & \centering 6.45 $\pm$ 0.28 \arraybackslash \\
            ReLyMe (in SongMASS) & \centering 6.57 $\pm$ 0.38 & \centering 6.33 $\pm$ 0.38 & \centering 6.39 $\pm$ 0.38 & \centering 6.46 $\pm$ 0.44 & \centering 6.41 $\pm$ 0.34 & \centering 6.39 $\pm$ 0.45 & \centering 6.51 $\pm$ 0.37 \arraybackslash \\
            ReLyMe (in TeleMelody) & \centering 7.12 $\pm$ 0.30 & \centering 7.01 $\pm$ 0.36 & \centering 7.06 $\pm$ 0.36 & \centering 6.91 $\pm$ 0.30 & \centering 6.93 $\pm$ 0.27 & \centering 6.81 $\pm$ 0.35 & \centering 6.95 $\pm$ 0.31 \arraybackslash \\
            \midrule
            \myname{}$_{\text{small}}$ & \centering 7.28 $\pm$ 0.32 & \centering 7.47 $\pm$ 0.29 & \centering 7.41 $\pm$ 0.41 & \centering 7.48 $\pm$ 0.33 & \centering 7.63 $\pm$ 0.27 & \centering7.57 $\pm$ 0.47 & \centering 7.60 $\pm$ 0.37 \arraybackslash \\
            \myname{}$_{\text{base}}$ & \centering \textbf{7.54 $\bm{\pm}$ 0.28} & \centering \textbf{7.78 $\bm{\pm}$ 0.38} & \centering \textbf{7.74 $\bm{\pm}$ 0.34} & \centering \textbf{7.66 $\bm{\pm}$ 0.25} & \centering \textbf{7.75 $\bm{\pm}$ 0.28} & \centering \textbf{7.83 $\bm{\pm}$ 0.39} & \centering \textbf{7.79 $\bm{\pm}$ 0.31} \arraybackslash \\
            \bottomrule
        \end{tabular*}
        \caption{Subjective results of \myname{} and baseline systems. Each score is calculated with 95\% confidence intervals.}
	\label{tab:subjective_results}
\end{table*}

\subsection{Evaluation Metrics}

In this subsection, we present a series of objective and subjective metrics used to evaluate the performance of \myname{} for lyric-to-melody generation.

\subsubsection{Objective Metrics}
\
\newline
We consider the following objective metrics to evaluate lyric-melody harmony by measuring the similarity between generated and ground-truth melodies \cite{zhang2022relyme,sheng2021songmass,ju2021telemelody}, in terms of pitch, duration, Inter Onset Interval (IOI)~\cite{yang2020evaluation}, and the overall melody sequence. Besides, we propose alignment distribution similarity to measure the consistency of word-note alignments across the generated and ground-truth melodies. To reduce the effects of stochastic sampling, we conduct each experiment on the test set 10 times.

\begin{itemize}
  \item \textbf{Pitch Distribution Similarity ($D_P$):} we compute the average Overlapped Area \cite{ren2020popmag} of the Pitch Class Histogram (PCH) distribution to evaluate the overall tonal similarity between the generated and ground-truth melodies.
  \item \textbf{Duration Distribution Similarity ($D_D$):} we quantize the note duration into 69 classes corresponding to 69 duration attributes in Table~\ref{tab:attributes}, and compute the average Overlapped Area of the duration distribution to evaluate the overall temporal similarity between the generated and ground-truth melodies.
  \item \textbf{IOI Distribution Similarity ($D_{IOI}$):} IOI is the time interval between two note onsets. We compute the average Overlapped Area of the IOI distribution to evaluate the overall rhythmic pattern similarity between the generated and ground-truth melodies.
  \item \textbf{Melody Distance ($MD$):} we convert the note sequence into a time series of pitches based on their durations, with a granularity of 10 (1/192 whole note), and perform pitch normalization by subtracting the average pitch of the entire sequence from each pitch. We use dynamic time warping \cite{berndt1994using} to measure the Euclidean Distance between the generated and ground-truth time series.
  \item \textbf{Alignment Distribution Similarity ($D_A$):} to evaluate the word-note alignment, we extract the alignment ($N$) distribution histogram for both generated and ground-truth songs based on the correspondence of one word to $N$ notes, and compute the average Overlapped Area.
\end{itemize}

\subsubsection{Subjective Metrics}
\
\newline
For subjective evaluation, we conduct a human listening test and compare \myname{} with the original and ReLyMe-equipped SongMASS and TeleMelody. We apply each model to generate 15 samples randomly and invite 10 participants, where 6 of them can understand basic music theory, to evaluate these samples. Specifically, we require all participants to score each sample using a ten-point scale (1 for lowest and 10 for highest) from two aspects: 1) the quality of the generated melody; 2) the quality of the overall sample, considering both the melody and corresponding lyrics. Detailed subjective metrics on these two aspects are described below. We utilize ACE Studio~\footnote{https://www.acestudio.ai} to synthesize the singing voice from lyrics and melodies, and provide participants with the lyrics, melodies and singing voice.

\noindent{\textbf{For the generated melody:}}
\begin{itemize}
    \item \textbf{Richness:} does the melody have rich and creative content?
    \item \textbf{Consistency:} does the melody sound smooth and have clear direction?
    \item \textbf{Listenability:} the overall listenability of the melody.
\end{itemize}

\noindent{\textbf{For the overall sample:}}
\begin{itemize}
    \item \textbf{Rhythmicity:} are the lyrics and melody consistent in terms of beat patterns and rests?
    \item \textbf{Structure:} do the melody and lyrics share similar structural patterns, such as reasonable repetition and variation?
    \item \textbf{Singability:} is the melody well matched with the lyrics in terms of quantity, and do the lyrics sound natural with the melody?
    \item \textbf{Overall:} the overall quality of the sample.
\end{itemize}

\subsection{Main Results}

To verify the effectiveness of \myname{} in the alignment and harmony between lyrics and melody, we compare \myname{}$_{\text{small}}$ to the original and ReLyMe-equipped SongMASS and TeleMelody with same system configurations. The objective results are shown in Table~\ref{tab:objective_results}. It is evident that \myname{}$_{\text{small}}$ significantly surpasses the baseline models across all objective metrics. Specifically, $D_A$ indicates that \myname{}$_{\text{small}}$ outperforms in lyric-melody alignment, while $D_P$, $D_D$, $D_{IOI}$ and $MD$ suggest that \myname{}$_{\text{small}}$ is the most capable of ensuring the harmony between lyrics and melodies. 
Table~\ref{tab:subjective_results} shows the subjective results, from which we can see that for melody itself, \myname{}$_{\text{small}}$ can generate diverse and consistent melodies. For the overall song, \myname{}$_{\text{small}}$ not only ensures the rhythmic and structural consistency between lyrics and melody, but also achieves the best results in singability and overall performance.
Besides, \myname{}$_{\text{base}}$ achieves better results with a larger model and pre-training dataset in both objective and subjective evaluations, showing the scalability and capability of \myname{}. 

\subsection{Method Analysis}

In this subsection, we analyze the effects of the designed 2D alignment encoding, lyric-melody relationships, and multi-task pre-training by conducting experiments on the following nine settings of \myname{}$_{\text{small}}$ with same configurations. 
Table~\ref{tab:objective_results} presents the overall objective results of these settings.

\begin{itemize}
    \item \textbf{Scratch:} train from scratch on the full dataset, without pre-training nor 2D alignment encoding.
    \item \textbf{CLM:} pre-train using casual language modeling (CLM).
    \item \textbf{CLM -- 2D Alignment Encoding:} pre-train using CLM, without 2D alignment encoding.
    \item \textbf{Random:} pre-train using a random span sampling strategy~\cite{joshi2020spanbert}.
    \item \textbf{Harmonized N-gram:} pre-train using our proposed harmonized n-gram sampling strategy.
    \item \textbf{Harmonized N-gram -- SMR:} pre-train using our proposed harmonized n-gram sampling strategy, without syllable stress and melodic peak relationship (SMR).
    \item \textbf{Harmonized N-gram -- SRR:} pre-train using our proposed harmonized n-gram sampling strategy, without syllable stress and rhythm skeleton relationship (SRR).
    \item \textbf{Phrase:} pre-train using a random phrase sampling strategy.
    \item \textbf{Long:} pre-train using a single long span sampling strategy~\cite{du2021glm}.
\end{itemize}

\noindent{\textbf{Effectiveness of 2D Alignment Encoding. }}
To verify the effectiveness of 2D alignment encoding, we compare the performance of CLM-based \myname{}$_{\text{small}}$ with and without 2D alignment encoding. For settings without 2D alignment encoding, we assign notes to each word in lyrics, ensuring that the number of notes for each word equals to the number of vowels in the word. As shown in Table~\ref{tab:objective_results}, our proposed 2D alignment encoding achieve much better scores on $D_A$, due to its excellent ability to directly capture the alignment between lyrics and melodies.

\noindent{\textbf{Effectiveness of Lyric-Melody Relationships. }}
To verify the contribution of each relationship, we conduct experiments based on the harmonized n-gram setting, excluding each relationship separately. The results in Table~\ref{tab:objective_results} show that the syllable stress and melodic peak relationship mainly contributes to pitch harmony ($D_P$) between lyrics and melodies, while the syllable stress and rhythm skeleton relationship plays an important role in rhythm harmony ($D_{IOI}$ and $D_D$).

\noindent{\textbf{Effectiveness of Multi-Task Pre-Training. }}
To further explore the benefits of multi-task pre-training, we compare it with single-task pre-training: 1) Harmonized N-gram, 2) Phrase, and 3) Long. The results in Table~\ref{tab:objective_results} show that our proposed multi-task pre-training method achieves the highest performance among all settings, demonstrating its excellent modeling performance on lyric-to-melody generation.

\section{Conclusion}

In this paper, we propose \myname{}, a lyric-to-melody generation system that leverages 2D alignment encoding and multi-task pre-training to ensure the alignment and harmony between lyrics and melodies. We introduce a unified symbolic song representation for lyrics and melodies that contains generic, content-related, and alignment-related attributes, and 2D alignment encoding to capture accurate alignments between lyrics and melodies. We design a multi-task pre-training framework with hierarchical blank infilling objectives (n-gram, phrase, and long span), and integrate lyric-melody relationships into the extraction of harmonized n-grams to guarantee the harmony between lyrics and melodies. Both objective and subjective results indicate that our proposed \myname{} can generate high-quality melodies from lyrics with remarkable lyric-melody alignment and harmony. Furthermore, method analysis shows the effectiveness of the detailed designs in \myname{}. In the future, we plan to extend our research to include more languages, such as Chinese, and explore the application of \myname{} to other automatic music composition tasks, such as text-to-music generation and video-to-music generation.

\section*{Acknowledgments}
This research was supported by the National Key Research and Development Program of China (No. 2023YFF0904900), the National Natural Science Foundation of China (No. 62272409), Guangdong OPPO Mobile Telecommunications Corp., Ltd., and ACE Studio.

\bibliography{aaai25}

\end{document}